\documentclass[a4paper,aps,pra,twocolumn,floatfix,superscriptaddress,notitlepage,usenames,dvipsnames,svgnames,table,footinbib,longbibliography]{revtex4-1}
\pdfoutput=1
\usepackage[utf8]{inputenc}
\usepackage[T1]{fontenc}
\usepackage{graphicx}
\usepackage{grffile}
\usepackage{wrapfig}
\usepackage{rotating}
\usepackage[normalem]{ulem}
\usepackage{amsmath}
\usepackage{textcomp}
\usepackage{amssymb}
\usepackage{capt-of}

\usepackage{parskip}
\usepackage{mathrsfs}
\usepackage[margin= 0.75in]{geometry}
\usepackage[braket, qm]{qcircuit}

\usepackage[english]{babel}
\usepackage{letltxmacro}
\LetLtxMacro{\ORIGselectlanguage}{\selectlanguage}
\makeatletter
\DeclareRobustCommand{\selectlanguage}[1]{%
  \@ifundefined{alias@\string#1}
    {\ORIGselectlanguage{#1}}
    {\begingroup\edef\x{\endgroup
      \noexpand\ORIGselectlanguage{\@nameuse{alias@#1}}}\x}%
}
\newcommand{\definelanguagealias}[2]{%
  \@namedef{alias@#1}{#2}%
}
\makeatother
\definelanguagealias{en}{english}
\definelanguagealias{eng}{english}

\usepackage{bbm}
\usepackage{etoolbox}
\usepackage{tcolorbox} 
\usepackage{tikz}
\usepackage{amsthm}
\usepackage{amssymb}
\usetikzlibrary{arrows.meta}
\usepackage{mathtools}

\usepackage{bm}
\usepackage{dsfont}
\usepackage[font=small,labelfont=bf,justification=RaggedRight,format=plain]{caption}
\usepackage{subcaption}
\usepackage{enumerate}
\usepackage{hhline}

\usepackage{thmtools}
\usepackage{thm-restate}

\newtheorem{proposition}{Proposition}
\newtheorem{lemma}{Lemma}

\usepackage{microtype}
\usepackage{dsfont}
\usepackage{tensor}

\usepackage{physics}
\usepackage{mathrsfs}
\usepackage{mathtools}
\usepackage{fixltx2e}
\usepackage{relsize}

\DeclarePairedDelimiterX\phys[2]{\langle}{\rangle}{#1 \delimsize\vert\mathopen{} #2}

\theoremstyle{remark}

\definecolor{blue-violet}{rgb}{0.54, 0.17, 0.89}
\usepackage{hyperref}
\hypersetup{
 pdfauthor={Faidon Andreadakis, Namit Anand, and Paolo Zanardi},
 pdftitle={Coherence generation, symmetry algebras and Hilbert space fragmentation},
 pdflang={English},colorlinks=true,linkcolor=RubineRed,citecolor=blue-violet,urlcolor=Cerulean} 
\usepackage[capitalise]{cleveref}

\begin{document}

\title{Coherence generation, symmetry algebras and Hilbert space fragmentation}

\author{Faidon Andreadakis}
\email [e-mail: ]{fandread@usc.edu}
\affiliation{Department of Physics and Astronomy, and Center for Quantum Information Science and Technology, University of Southern California, Los Angeles, California 90089-0484, USA}

\author{Paolo Zanardi}
\email [e-mail: ]{zanardi@usc.edu}
\affiliation{Department of Physics and Astronomy, and Center for Quantum Information Science and Technology, University of Southern California, Los Angeles, California 90089-0484, USA}
\affiliation{Department of Mathematics, University of Southern California, , Los Angeles, California 90089-2532, USA}

\date{\today}

\begin{abstract}
Hilbert space fragmentation is a novel type of ergodicity breaking in closed quantum systems. Recently, an algebraic approach was utilized to provide a definition of Hilbert space fragmentation characterizing \emph{families} of Hamiltonian systems based on their (generalized) symmetries. In this paper, we reveal a simple connection between the aforementioned classification of physical systems and their coherence generation properties, quantified by the coherence generating power (CGP). The maximum CGP (in the basis associated to the algebra of each family of Hamiltonians) is exactly related to the number of independent Krylov subspaces $K$, which is precisely the characteristic used in the classification of the system. In order to gain further insight, we numerically simulate paradigmatic models with both ordinary symmetries and Hilbert space fragmentation, comparing the behavior of the CGP in each case with the system dimension. More generally, allowing the time evolution to be any unitary channel in a specified algebra, we show analytically that the scaling of the Haar averaged value of the CGP depends only on $K$. These results illustrate the intuitive relationship between coherence generation and symmetry algebras.
\end{abstract}
\maketitle
\section{Introduction} \label{secintro}
\par Non-equilibrium quantum dynamics of closed systems has been an extensively studied issue in recent years. Generic isolated quantum systems are expected to thermalize in the thermodynamic limit, in the sense that for any small subsystem the rest of the system acts as a bath, so that the expectation value of local observables coincides with that derived by statistical mechanical ensembles \cite{nandkishore_many-body_2015,gogolin_equilibration_2016}. A central role in the understanding of this phenomenon is played by the so-called eigenstate thermalization hypothesis (ETH) \cite{deutsch_quantum_1991,srednicki_chaos_1994,rigol_thermalization_2008}, according to which, under certain conditions, the energy eigenstates are conjectured to be thermal, i.e. the expectation values of observables with local support over the energy eigenstates coincide with those of a thermal ensemble with temperature related to the energy \cite{dalessio_quantum_2016}.
\par While ETH is expected to hold for generic systems \cite{rigol_thermalization_2008,rigol_quantum_2010,ikeda_eigenstate_2011,dubey_approach_2012,
steinigeweg_eigenstate_2013,kim_testing_2014,beugeling_finite-size_2014,steinigeweg_pushing_2014,muller_thermalization_2015,mondaini_eigenstate_2016}, several ETH-violating classes of models are known. Integrable systems and many-body localized (MBL) systems are the two prime examples, where an extensive number of conserved quantities prevent the system from thermalizing. The conserved quantities are a result of symmetries readily encoded in the Hamiltonian (in the integrable systems) or from emergent symmetries created by strong disorder (in the MBL case). More novel types of disorder-free ergodicity breaking models were recently observed and studied, dubbed quantum many-body scars (QMBS) \cite{bernien_probing_2017,shiraishi_systematic_2017,turner_weak_2018,turner_quantum_2018,
moudgalya_exact_2018,lin_exact_2019,ok_topological_2019,pai_dynamical_2019,
mark_unified_2020,hudomal_quantum_2020,iadecola_nonergodic_2020,iadecola_quantum_2020,
mcclarty_disorder-free_2020,langlett_hilbert_2021,papaefstathiou_disorder-free_2020,
desaules_proposal_2021,moudgalya_large_2020,banerjee_quantum_2021,lee_exact_2020,
lin_quantum_2020,schecter_weak_2019,mark_ensurematheta-pairing_2020,pakrouski_many-body_2020,moudgalya_ensurematheta-pairing_2020,moudgalya_exhaustive_2022} and Hilbert space fragmentation \cite{khemani_localization_2020,sala_ergodicity-breaking_2020,yang_hilbert-space_2020,rakovszky_statistical_2020,herviou_many-body_2021,hahn_information_2021,langlett_hilbert_2021,moudgalya_thermalization_2021,
de_tomasi_dynamics_2019,lee_frustration-induced_2021,karpov_disorder-free_2021,zhang_hilbert_2022,khudorozhkov_hilbert_2022,moudgalya_hilbert_2022}. QMBS correspond to a weak violation of ETH, in which only a small set of eigenstates in the bulk of the spectrum are non-thermal. Emergence of scar dynamics has been linked with the existence of dynamical symmetries \cite{medenjak_isolated_2020,pakrouski_many-body_2020,odea_tunnels_2020,ren_quasisymmetry_2021}, that allow the formation of towers of scar eigenstates inside an invariant subspace. Hilbert space fragmentation corresponds to the broader phenomenon of "shattering" of the Hilbert space in exponentially many dynamically disconnected subspaces (referred to as Krylov subspaces) with no obvious associated conserved quantities. In addition to weak violation, such systems can exhibit a strong violation of ETH, where a nonzero measure subset of all the eigenstates is non-thermal.
\par Recently, an algebraic framework was utilized to analyze the phenomenon of Hilbert space fragmentation for families of Hamiltonians \cite{moudgalya_hilbert_2022}, establishing the role of non-conventional symmetries. Such an approach reveals the central role of the symmetry algebra in the classification of the system, and is applicable to conventional models as well \cite{moudgalya_symmetries_2022}. Another inherent advantage is that the properties exhibited by the families of Hamiltonians are free from fine-tuning. In general, systems exhibiting novel ergodic behavior (such as QMBS and Hilbert space fragmentation) belong to special classes of Hamiltonians \cite{moudgalya_hilbert_2022} that are of great interest in attempts to fully describe the behavior of non-equilibrium closed quantum systems \cite{sala_ergodicity-breaking_2020,moudgalya_exact_2018,moudgalya_entanglement_2018,iadecola_exact_2019,iadecola_quantum_2019,ok_topological_2019}.
\par The ergodic properties of a system are correlated with information-theoretic signatures, such as eigenstate entanglement \cite{khare_localized_2020,noauthor_phys_nodate,papic_weak_2022} and information scrambling \cite{sun_quantum_2021,yuan_quantum_2022,turner_weak_2018}. In this paper, we establish a connection between the classification framework introduced in \cite{moudgalya_hilbert_2022} and coherence, quantified by the coherence generating power (CGP) \cite{zanardi_measures_2017,zanardi_coherence-generating_2017,zanardi_quantum_2018,styliaris_coherence-generating_2018}. The CGP bound established in this paper becomes particularly relevant for the aforementioned special classes of Hamiltonians exhibiting Hilbert space fragmentation.
\par This paper is structured as follows. In \cref{sec_bound}, we present an analytical upper bound of the CGP of \emph{families} of Hamiltonian evolutions, with respect to a basis induced by the algebra of each family; this bound is fully described by the aforementioned classification of the families of Hamiltonians. We, also, provide numerical results of systems with both ordinary symmetries and Hilbert space fragmentation and compare the behavior of the CGP as we increase the system size. In \cref{sec_haar}, we allow the evolution to be any unitary channel in a prescribed algebra and show that the Haar average value of the CGP is typical and its scaling with the system size is fully described by the classification of unitary evolutions in the algebra. In \cref{sec_conclusion}, we conclude with a brief discussion of the results. The detailed proofs of the analytical results are included in the Supplemental Material \ref{appendix}.

\section{Coherence generating power for classes of models} \label{sec_bound}
\subsection{Preliminaries} \label{subsec_prelim}
Let $\mathcal{H}\cong\mathbb{C}^d$ be a finite $d$-dimensional Hilbert space and denote the space of linear operators acting on it as $\mathcal{L}(\mathcal{H})$. The key mathematical structures of interest are $*$-closed untial algebras of observables $\mathcal{A}$, alongside their commutants,
\begin{equation*}
\mathcal{A}^\prime = \{ Y \in \mathcal{L} (\mathcal{H} ) \; \vert \; [X,Y] =0 \; \forall \; X \in \mathcal{A} \}
\end{equation*}
Due to the double commutant theorem these algebras always come in pairs, so as $(\mathcal{A}^\prime )^\prime = \mathcal{A}$ \cite{davidson_c-algebras_1996}.
\par Letting $\mathcal{Z}(\mathcal{A} ) := \mathcal{A} \cap \mathcal{A}^\prime$ denote the center of $\mathcal{A}$ with $d_\mathcal{Z} := \dim \mathcal{Z} (\mathcal{A} )$, a fundamental structure theorem for $C^*$-algebras implies a decomposition of the Hilbert space of the form \cite{davidson_c-algebras_1996}
\begin{equation} \label{eq_decomposition}
\begin{split}
&\mathcal{H} \cong \oplus_{J=1}^{d_{\mathcal{Z}}} \,  \mathbb{C}^{n_J} \otimes \mathbb{C}^{d_J}, \\
&\mathcal{A} \cong \oplus_{J=1}^{d_{\mathcal{Z}}} \,  \mathds{1}_{n_J} \otimes \mathcal{L}(\mathbb{C}^{d_J}), \\
&\mathcal{A}^\prime \cong \oplus_{J=1}^{d_Z} \, \mathcal{L}(\mathbb{C}^{n_J}) \otimes \mathds{1}_{d_J}.
\end{split}
\end{equation}
Due to the above decomposition,
\begin{equation*}
\begin{split}
&\dim\mathcal{H}\equiv d = \sum_{J=1}^{d_{\mathcal{Z}}} n_J d_J, \\
&\dim\mathcal{A} = \sum_{J=1}^{d_{\mathcal{Z}}} d_J^2 =: d(\mathcal{A}), \\
&\dim\mathcal{A}^\prime = \sum_{J=1}^{d_{\mathcal{Z}}} n_J^2 =: d(\mathcal{A}^\prime).
\end{split}
\end{equation*}
In general the algebra $\mathcal{A}^\prime$ is non-Abelian and we will denote as $\mathcal{M} \subset \mathcal{A}^\prime$ the (possibly not unique) maximal Abelian subalgebra of the commutant with dimension $K:= \dim \mathcal{M}=\sum_{J=1}^{d_\mathcal{Z}} n_J$. Note that $K$ is exactly the number of linearly independent \emph{common} invariant subspaces of all operators in $\mathcal{A}$.
\par Given a basis $B= \{ \ket{i} \}_{i=1}^d$ of $\mathcal{H}$, one can always express a pure state as a linear superposition of the states $\ket{i}$. This fact is experimentally materialized as what is called quantum coherence \cite{glauber_coherent_1963}. In general, one defines $B$-incoherent states $\rho$ ($\rho \geq 0 , \Tr \rho = 1$) as states diagonal in $B$ (coherent states are all states that are not incoherent). Given a unitary operator $U \in \mathcal{L}(\mathcal{H} )$ a measure of its coherence generating power (CGP) with respect to the basis $B$ is given by \cite{zanardi_coherence-generating_2017}
\begin{equation} \label{eq_cgp_def}
C_B (U) = 1-\frac{1}{d} \sum_{i,j=1}^d \lvert \mel{i}{U}{j} \rvert^4
\end{equation}
Note that the above quantity coincides with a measure of information scrambling of $B$-diagonal ("incoherent") degrees of freedom \cite{zanardi_quantum_nodate,andreadakis_scrambling_2022}, which aligns with its ability of quantifying the deviation of incoherent states from the incoherent space under evolution with $U$.
\par Ref. \cite{moudgalya_hilbert_2022} introduced an algebra based classification of multi-site Hamiltonian models. Given an algebra $\mathcal{A}$ that can be generated by local operators $h_j$, one considers the family of Hamiltonians
\begin{equation} \label{eq_family}
H=\sum_j J_j \; h_j
\end{equation}
where $J_j$ are arbitrary coupling constants. Then, the symmetries of the system are characterized by the scaling of $K$ with the system size. For example, for 1D systems of size $L$, if $\log K \sim L$, then the \emph{family} of Hamiltonians in \cref{eq_family} possesses an exponentially large number of dynamically disconnected subspaces, which serves as the definition of Hilbert space fragmentation (see \cref{table_classification}). The models exhibiting Hilbert space fragmentation were shown to exhibit nonlocal conserved quantities, dubbed as statistically localized integrals of motion (SLIOMs) \cite{rakovszky_statistical_2020,moudgalya_hilbert_2022}.
\begin{table}
\centering
\begin{tabular}{|l|c|}
\hline
$\log K$\hspace{10pt}&Class representative\\
\hhline{|=|=|}
$\sim\mathcal{O}(1)$\hspace{10pt}&Discrete global symmetry\\
$\sim \log L$\hspace{10pt}&Continuous global symmetry\\
$\sim L$\hspace{10pt}&Fragmentation\\
\hline
\end{tabular}
\caption{Classification of families Hamiltonians of the form \cref{eq_family} based on the scaling of the dimension $K$ of the maximally abelian subalgebra of the commutant $\mathcal{A}^\prime$ with system size $L$ in 1D.\cite{moudgalya_hilbert_2022}}
\label{table_classification}
\end{table}
\subsection{Coherence generating power bound} \label{subsec_bound}
We consider an algebra $\mathcal{A}=\langle h_j\rangle$ and families of Hamiltonians of the form \cref{eq_family}. Clearly, $H \in \mathcal{A}$ and due to \cref{eq_decomposition}
\begin{equation} \label{eq_unitary}
U:= \exp(-it \, H) = \oplus_{J=1}^{d_\mathcal{Z}} \mathds{1}_{n_J} \otimes U_{J}
\end{equation}
We also consider a basis $B_\mathcal{A} = \{ \ket{\phi_J} \otimes \ket{\psi_J} \; \vert \; J=1,\dots,d_\mathcal{Z}, \; \phi_J = 1,\dots ,n_J, \; \psi_J = 1, \dots , d_J \}$, where $\ket{\phi_J}$ and $\ket{\psi_J}$ span $\mathbb{C}^{n_J}$ and $\mathbb{C}^{d_J}$ of the decomposition \cref{eq_decomposition}. We can, then, show that
\begin{proposition} \label{prop_cgp}
\begin{enumerate}[i)]
\item {The CGP of $U=\exp(-it \, H)\in \mathcal{A}$ in a basis $B_\mathcal{A}$ induced by the algebra decomposition is
\begin{equation} \label{eq_cgp}
\begin{split}
C_{B_\mathcal{A}} (U) &= 1- \frac{1}{d} \sum_{J,\psi_J , \psi^\prime_J} n_J \; \lvert \mel{\psi_J}{U_J}{\psi^\prime_J} \rvert^4\\
&=: 1-\frac{1}{d} f_{B_\mathcal{A}}(U)
\end{split}
\end{equation}}
\item {The maximum value of the CGP is 
\begin{equation} \label{eq_max}
\max_{U \in \mathcal{A}} C_{B_\mathcal{A}} (U) = 1-\frac{1}{d} K
\end{equation}
and is achieved when $\lvert \mel{\psi_J}{U_J}{\psi^\prime_J} \rvert = d_J^{-1/2}$ for all $J$.}
\end{enumerate}
\end{proposition}
We note that if $U$ is any unitary operator in $\mathcal{L}(\mathcal{H})$, then the maximum CGP is $C_{max}=1-\frac{1}{d}$ \cite{zanardi_coherence-generating_2017,zanardi_quantum_nodate,andreadakis_scrambling_2022}. The difference in the bound of \cref{eq_max} is an extra factor of $K$, which is exactly the number of independent \emph{common} invariant subspaces for all evolutions generated by the family of Hamiltonians in \cref{eq_family}. The scaling of $K$ with the system dimension provides a classification of this family (see e.g. \cref{table_classification}), thus from \cref{eq_max} the scaling of the maximum CGP of the entire family of Hamiltonians is also dependent \emph{solely} on this classification as well. This observation constitutes an exact, analytical implementation of the intuitive connection between (generalized) system symmetries and coherence generation. Note that the Hilbert space decomposition \cref{eq_decomposition} is a basic element for the protection of quantum coherence in information processes, e.g. by decoherence-free subspaces and subsystems \cite{noauthor_phys_nodate-1,zanardi_dissipation_1998,lidar_decoherence-free_1998,noauthor_phys_nodate-2}. There, the algebra $\mathcal{A}$ is generated by the system operators of the interaction Hamiltonian and $K$ is simply the total dimension of decoherence-free subsystems $\mathbb{C}^{n_J}$.

\subsection{Quantum spin-chain models}\label{spin_chain_sect}
\begin{figure*}
\centering
\begin{subfigure}{.5\textwidth}
  \centering
  \includegraphics[width=1\linewidth]{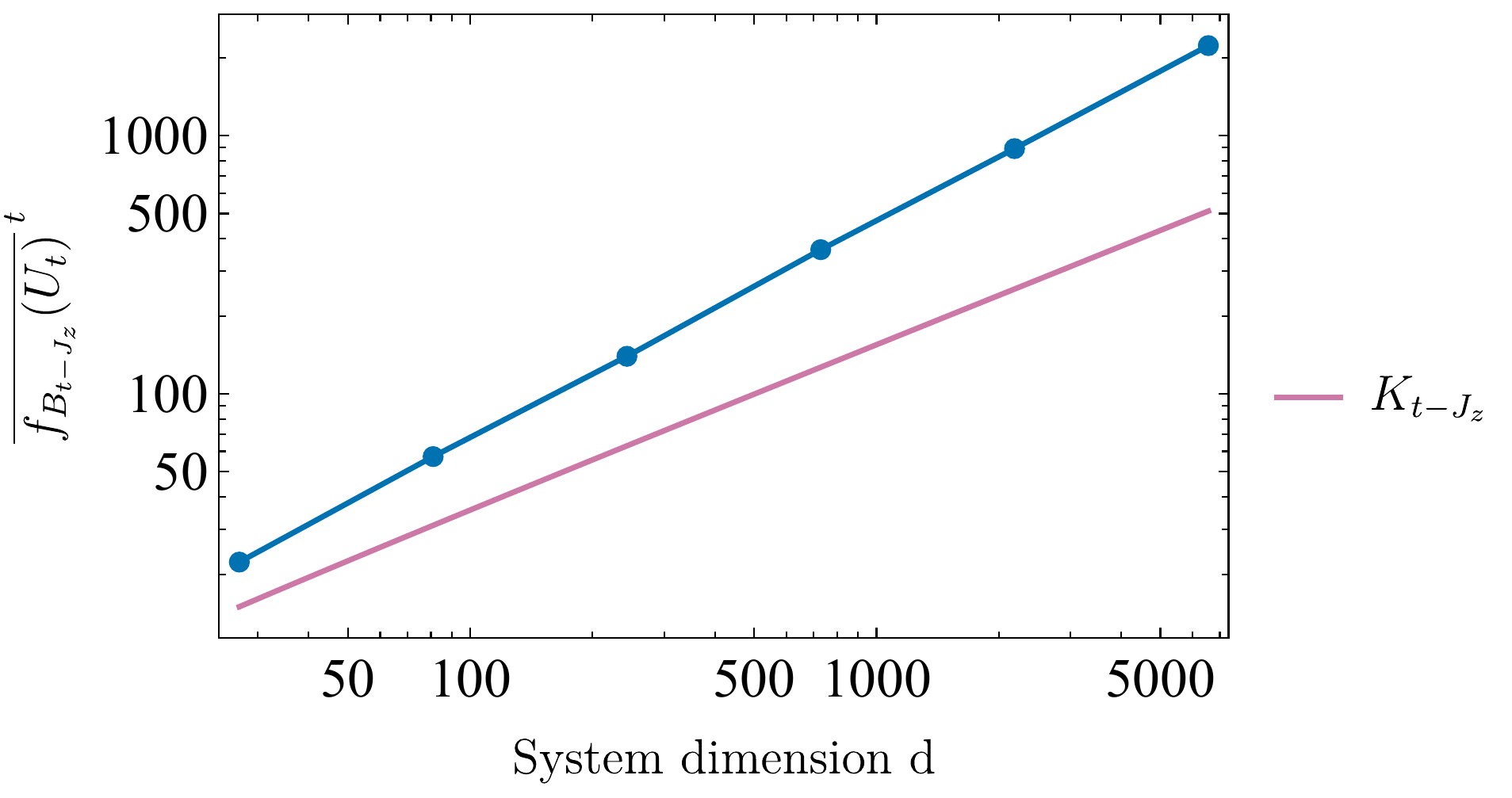}
 \caption{}\label{fig_tJz}
\end{subfigure}%
\begin{subfigure}{.5\textwidth}
  \centering
  \includegraphics[width=1\linewidth]{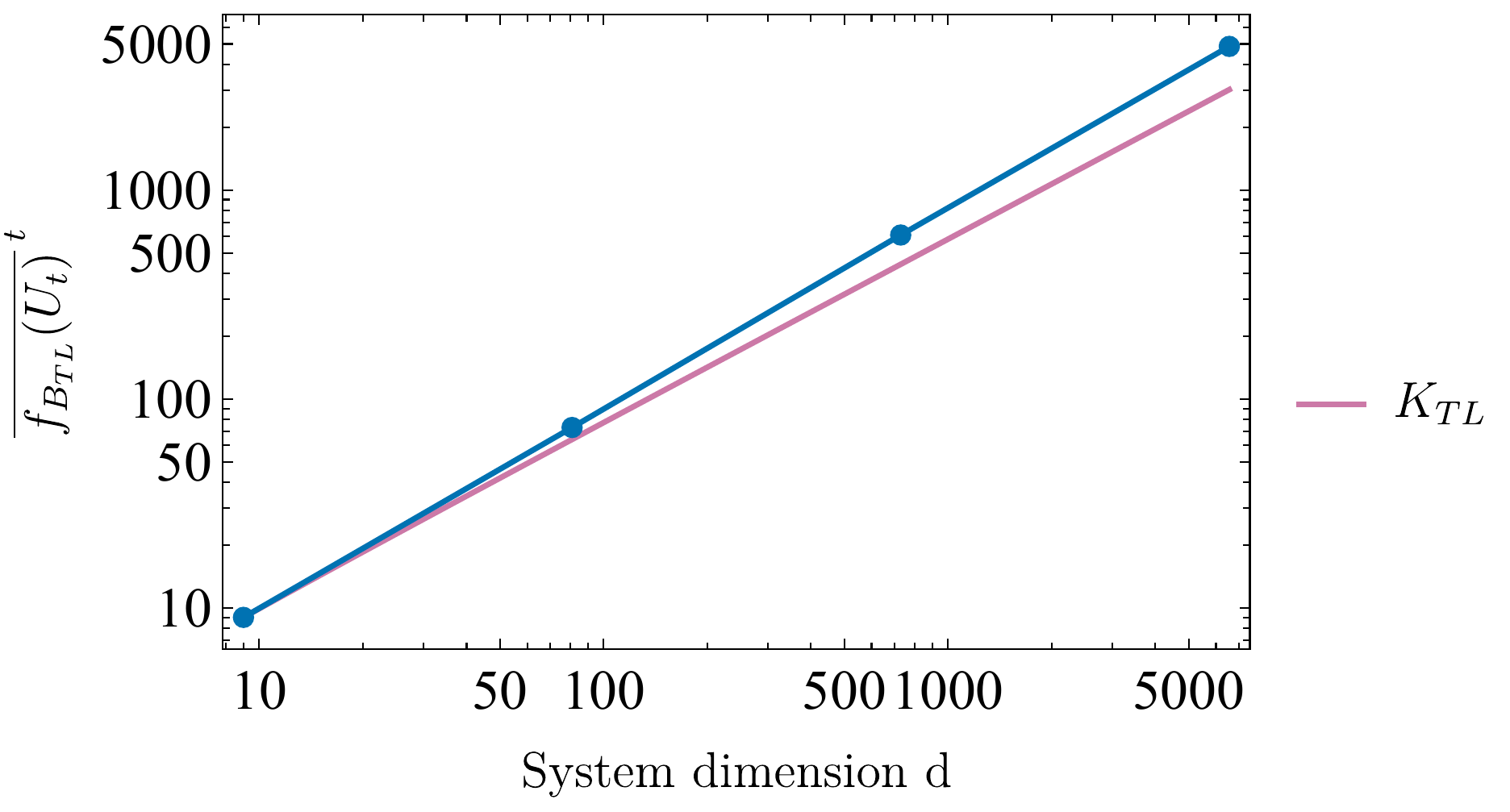}
  \caption{}\label{fig_TL}
\end{subfigure}
\begin{subfigure}{.55\textwidth}
  \centering
  \includegraphics[width=1\linewidth]{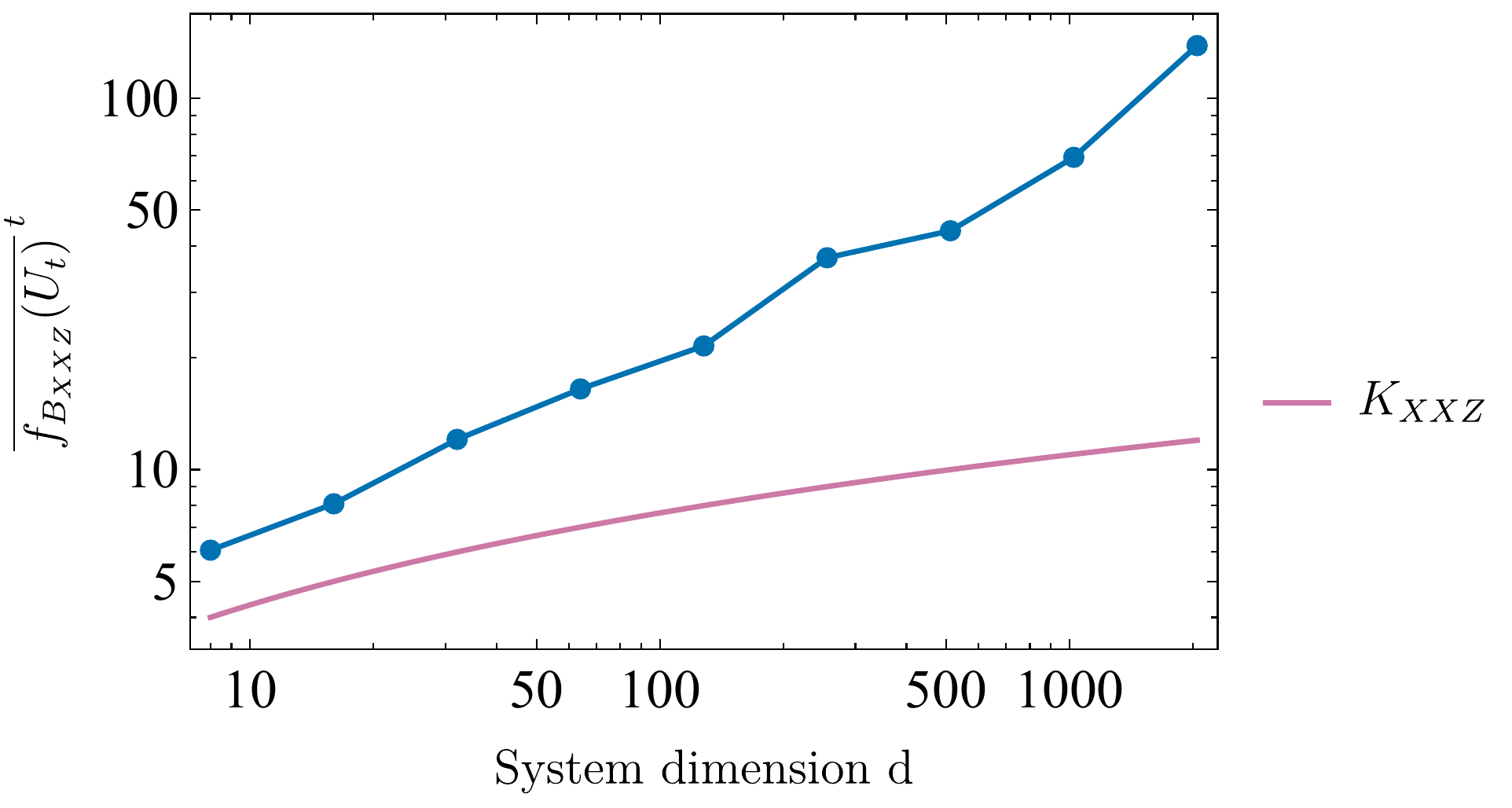}
 \caption{}\label{fig_XXZ}
\end{subfigure}
\caption{ Log-log plots of long-time average of $f_{B_\mathcal{A}} (U)=1-C_{B_\mathcal{A}}(U)$ (connected dotted lines) in comparison to the bound $K$ (continuous lines) for: (a) the $t-J_z$ model; $\overline{f_{B_{t-J_z}} ({U_t})}^t \sim d^{0.83}$ scales exponentially with the system size. The bound set by the number of independendent Krylov subspaces scales as $d^{\log_32}$, (b) the $TL$ model; $\overline{f_{B_{TL}} ({U_t})}^t \sim d^{0.95}$ scales exponentially with the system size. The scaling is very close to that set by the bound $d^{\log_3q}$ (see \cref{dimensionality_details}), $\log_3q\simeq 0.88$, which can be understood in terms of the existence of Krylov-restricted thermalization in some of the large fragments of the $TL$ model \cite{moudgalya_thermalization_2021}, (c) the $XXZ$ model; the behavior of $\overline{f_{B_{XXZ}} ({U_t})}^t$ greatly deviates from the bound with the system size. Integrable systems (even with a random choice of couplings) are not expected to showcase generic properties, even after resolving for symmetries.}
\label{fig_f}
\end{figure*}

\begin{figure}
\centering
\includegraphics[width=1\linewidth]{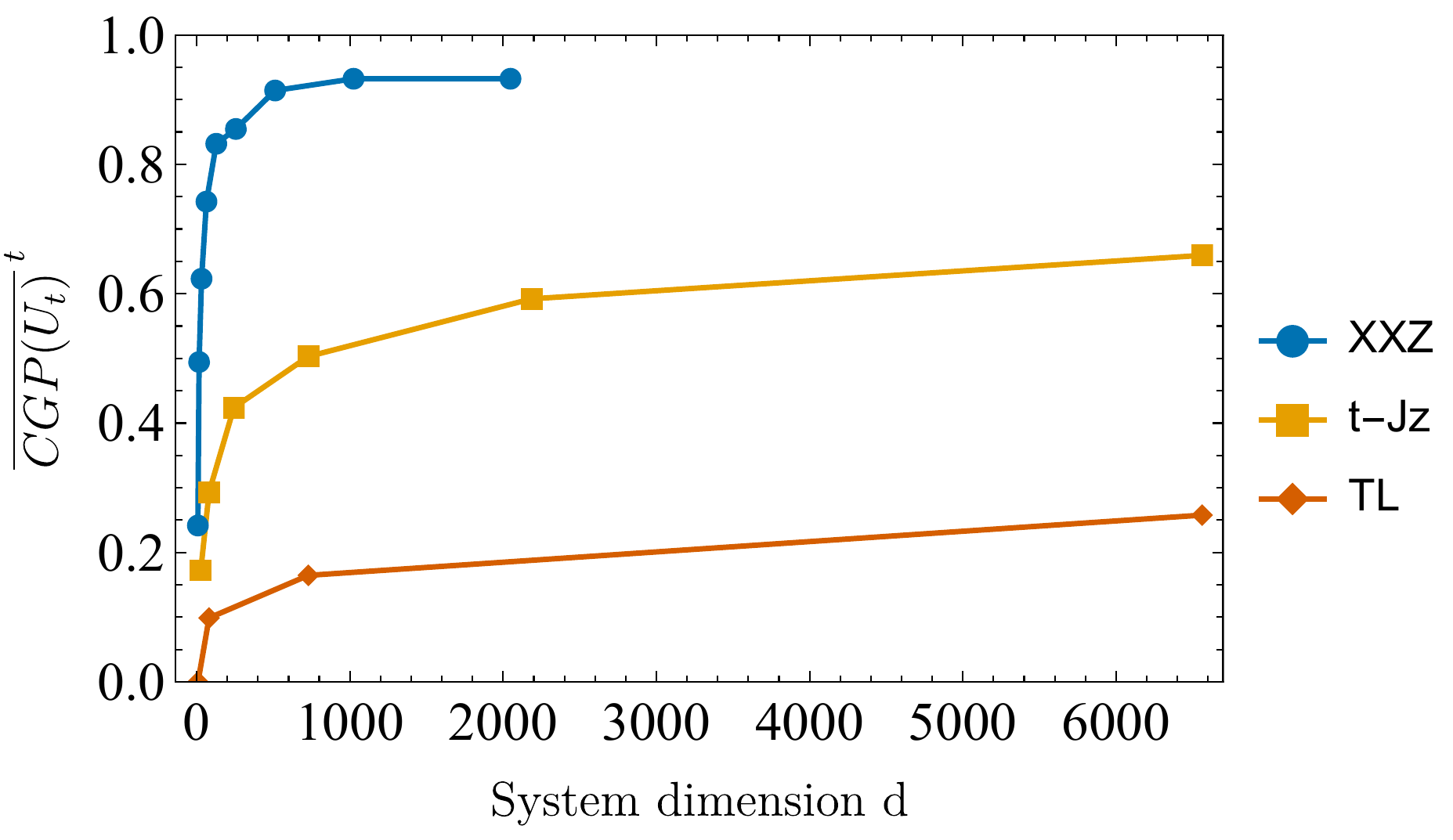}
\caption{Comparison of the scaling of the CGP for the $XXZ$, $t-J_z$ and $TL$ models. We observe that for finite dimensions, the CGP of fragmented models is heavily constraint by the bound set in \cref{eq_max}, leading to less mixing of different parts of the Hilbert space under evolution}
\label{fig_cgp}
\end{figure}
Note that the result in \cref{prop_cgp} only depends on the fact that $U\in \mathcal{A}$ and not on the fact that $U$ is generated by a Hamiltonian of the form \cref{eq_family}. As so, if we choose randomly a Hamiltonian from \cref{eq_family}, the resulting evolution need not be typical. In order to evaluate the behavior of the CGP in such a case, we numerically simulate the evolution of spin-chain models in 1D with $L$ sites: (i) the spin-1/2 XXZ model with on-site magnetic field, (ii) the spin-1 Temperley-Lieb (TL) model, and (iii) the fermionic $t-J_z$ model.
\begin{equation} \label{eq_hamiltonians}
\begin{split}
&H_{XXZ}=\sum_j \left( J_j^\perp \left( \sigma_j^x \sigma_{j+1}^x + \sigma_j^y \sigma_{j+1}^y \right)+ \right. \\
& \left. \hspace{65pt}+ J_j^z \sigma_j^z \sigma_{j+1}^z + h_j \sigma_j^z \right) \\
&H_{TL} = \sum_j J_j e_{j,j+1}, \text{ where }e_{j,k}= 3(\ket{\psi_{sing}}\bra{\psi_{sing}})_{j,k}\\
&\hspace{80pt} \text{ and } \ket{\psi_{sing}}_{j,k}= \frac{1}{\sqrt{3}}\sum_{\alpha\in \{0,1,2\}} \ket{\alpha \alpha}_{j,k}\\
&H_{t-J_z}=\sum_j \left( -t_{j,j+1} \sum_{\sigma \in \{ \uparrow , \downarrow \} } \left( \tilde{c}_{j,\sigma} \tilde{c}^\dagger_{j+1,\sigma} + h.c. \right) + \right. \\
&\left. \hspace{80pt}+ J_{j,j+1}^z S_j^z S_{j+1}^z + h_j S_j^z + g_j \left(S_j^z \right)^2 \right), \\
&\hspace{50pt}\text{where }S_j^z := \tilde{c}_{j,\uparrow}^\dagger \tilde{c}_{j,\uparrow} - \tilde{c}_{j,\downarrow}^\dagger \tilde{c}_{j,\downarrow}\\
& \hspace{50pt} \text{and }\tilde{c}_{j,\sigma} := c_{j,\sigma} \left( 1-c_{j,-\sigma}^\dagger c_{j,-\sigma} \right)
\end{split}
\end{equation}
\par Here, $\sigma_j^\alpha, \alpha \in {x,y,z}$ denote the Pauli matrices. The XXZ model has shown to be integrable by Bethe ansatz \cite{takahashi_thermodynamics_1999} and possesses a conventional $U(1)$ symmetry with the associated conserved quantity $\sigma_{tot}^z=\sum_j \sigma_j^z$. The sectors of \cref{eq_decomposition} are labeled by this conserved quantity ($J=L/2 - \sigma_{tot}^z= 0,\dots ,L$) and the dimensions of the irreducible representations of $\mathcal{A}$ and $\mathcal{A}^\prime$ are $d_J= {L \choose J}$ and $n_J=1$ respectively \cite{moudgalya_hilbert_2022}. For a given $J$ there is one common Krylov subspace (since $n_J=1$) and is spanned by the $z-basis$ product states with $\sigma_{tot}^z = L/2 - J$). Note that $K_{XXZ}=L+1$, which scales linearly with the system size.
\par The TL model is an example of Hilbert space fragmentation. The dynamically disconnected Krylov subspaces are understood by the use of a basis of dots and non-crossing dimers \cite{read_enlarged_2007,batchelor_spin-s_1990,moudgalya_hilbert_2022}. Each basis state consists of a pattern of dimers that connect two sites and represent the state $\ket{\psi_{sing}}_{j,k}$. The rest of the sites (that are not connected to either end of a dimer) make up the dots and the state $\ket{\psi_{dots}}$ is chosen such that it is annihilated by all projectors $\ket{\psi_{dim}}_{m,m+1} \bra{\psi_{dim}}_{m,m+1}$ (where $m$ labels all dots while excluding the dimers). The algebra formed by the $L-1$ generators $e_{j,j+1}$ is the Temperley-Lieb algebra $TL_{L}(q)$, where $q+q^{-1}=3$ (which is the number of local degrees of freedom in a spin-1 model)\cite{moudgalya_hilbert_2022}. It has been shown that, for even $L$, the sectors of the decomposition \cref{eq_decomposition} are labeled by $J=0,\dots,L/2$ with corresponding dimensions 
\begin{equation*}
n_J=[2J+1]_q, \; d_J={L \choose L/2+J} - {L \choose L/2+J+1}
\end{equation*}
where $[n]_q := (q^n-q^{-n})/(q-q^{-1})$ is a $q$-deformed integer \cite{read_enlarged_2007}. It is, then, straightforward to show that $K=\frac{q^{-L} \left(q^{L+2}-1\right)^2}{\left( q^2 -1 \right)^2}$, which scales exponentially with the system-size.
\par The $t-J_z$ also falls into the category of Hilbert space fragmentation. The Hamiltonian acts effectively in the Hilbert space with no double occupancy and it has been shown that all operators in the commutant algebra are diagonal in the product state basis\cite{moudgalya_hilbert_2022}. The sectors are characterized by the pattern of spins $\uparrow$ and $\downarrow$, which remains invariant under the action of the Hamiltonian\cite{rakovszky_statistical_2020}. For open boundary conditions (OBC) there are $2^{L+1}-1$ such sectors (all with $n_J=1$), so $K=2^{L+1}-1$ which scales exponentially with $L$\cite{moudgalya_hilbert_2022}.
\par We simulate exact dynamics for the above systems and for different system-sizes with the coupling constants randomly drawn from $[0,1]$ (each time we increase the system-size we add one more set of randomly chosen couplings to the previously drawn ones). The choice of the coupling constants sets the energy scale of the Hamiltonians, and in turn the timescale of the dynamics. For each model we construct the relevant bases described above and compute the long-time average of the CGP of the unitary evolution. In \cref{fig_tJz}, \ref{fig_TL} we observe that for the fragmented models the long-time average of the quantity $f_{B_\mathcal{A}} (U)$ (related to the CGP via \cref{eq_cgp}) has a similar behavior with the number of Krylov subspaces $K$. Specifically, we find that (see also \cref{dimensionality_details})
\begin{equation*}
\begin{split}
&\overline{f_{B_{t-J_z}} ({U_t})}^t \sim d^{0.83} \text{(compared to }K_{t-J_z} \sim d^{\log_32 })\\
&\overline{f_{B_{TL}} ({U_t})}^t \sim d^{0.95}\text{ (compared to }K_{TL} \sim d^{\log_3q}, \\
&\hspace{100pt} \; q=(3+\sqrt{5})/2\simeq 2.62\text{)}
\end{split}
\end{equation*}
Fragmented systems can be further characterized into strongly or weakly fragmented depending on the size $\max_J d_J$ of the biggest Krylov subspace; specifically for strongly (weakly) fragmented systems, $\max_J d_J / d \rightarrow 0$ ($\max_J d_J / d \rightarrow 1$) as $d \rightarrow \infty$ \cite{sala_ergodicity-breaking_2020,moudgalya_hilbert_2022}. When the size of the fragments is large enough (in the thermodynamic limit) and the Hamiltonian is non-integrable in these fragments, signatures of thermalization within the fragment can be observed, a phenomenon referred to as Krylov-restricted thermalization \cite{moudgalya_thermalization_2021}. In fact, the $TL$ model has been shown to exhibit thermalization in some of its exponentially large Krylov subspaces \cite{moudgalya_thermalization_2021}, which implies that the behavior of the CGP is expected to be closer to that of a generic model (see also \cref{sec_haar}). In contrast to the fragmented models, the behavior of the CGP of the integrable $XXZ$ model with the system dimension is far away from the bound \cref{eq_max} (\cref{fig_XXZ}). This aligns with the expectation that integrable models do not showcase features of generic evolutions, even after resolving for the symmetries. \cref{fig_cgp} emphasizes the vastly different CGP behaviors of the different classes of models simulated; for finite dimensions the bound \cref{eq_max} forces the CGP of the fragmented models to remain much lower than that of the integrable model. This shows that despite the fact that fragmented models are in general non-integrable, emergent generalized symmetries from kinetic constraints lead to a milder mixing of different parts of the Hilbert space, with possible physical importance for information processing tasks, e.g. protection of information via decoherence free subspaces \cite{zanardi_error_1997,zanardi_noiseless_1997,lidar_decoherence-free_1998} or utilization of (dynamical) localization for quantum memory implementations \cite{hahn_information_2021,smith_many-body_2016}.

\section{Haar averaged coherence generating power and system-size scaling} \label{sec_haar}
We will now consider the case where $U \in \mathcal{A}$ can be any unitary in the algebra. It is straightforward to notice that \cref{prop_cgp} continues to hold, since \cref{eq_unitary} holds for all unitaries in the algebra. A natural question we wish to investigate is, given an algebra $\mathcal{A}$, what the typical value of the CGP is and how it is related to the number of independent Krylov subspaces $K$. To do so, we average $C_{B_\mathcal{A}}(U)$ over the Haar measure of the subgroup of unitaries $U\in \mathcal{A}$.
\begin{proposition} \label{prop_haar}
Given an algebra $\mathcal{A}$, the Haar average of the CGP over the unitaries in the algebra is
\begin{equation} \label{eq_haar}
\overline{C_{B_\mathcal{A}}(U)}^{U \in \mathcal{A}} := {\mathlarger{\mathbb{E}}}_{U \in \mathcal{A}} \left[C_{B_\mathcal{A}} (U) \right] = 1- \frac{1}{d} \sum_J \frac{2d_J}{d_J+1} \; n_J
\end{equation}
\end{proposition}
\cref{eq_haar} provides an analytical expression for the Haar average value of the CGP in terms of the dimensions $n_J, \; d_J$ related to the decomposition \cref{eq_decomposition}. Since $1\leq d_J \leq d \; \forall J$  we can (loosely) bound the typical value as
\begin{equation}\label{eq_haar_bounds}
1-\frac{2}{d+1} \; K \leq \overline{C_{B_\mathcal{A}}(U)}^{U \in \mathcal{A}} \leq 1 - \frac{1}{d} \; K
\end{equation}
This shows, that the scaling of the average value with the system dimension depends exactly on the classification of the model in terms of the scaling of $K$ with the system size (\cref{subsec_prelim}). As seen in Proposition 4 of \cite{zanardi_coherence-generating_2017}, Levy's lemma implies that the Haar average is expected to be typical inside sufficiently large Krylov subspaces (see \cref{append_typicality}). This aligns with the observation that the weakly fragmented $TL$ model seems to have close to maximal CGP, as in that case there is a dominant Krylov subspace, such that $\max_J \frac{d_J}{d} \rightarrow 1$ in the thermodynamic limit.
\par Note that due to the double commutant theorem, $\mathcal{A}$ is completely determined by the commutant $\mathcal{A}^\prime$, which represents the set of symmetries imposed on the evolution. As so, the selection of a random unitary in $\mathcal{A}$ corresponds to a random unitary channel constrained by physically motivated symmetries.


\section{conclusion} \label{sec_conclusion}
\par In this paper we revealed a connection between a classification of families of Hamiltonian evolutions and their coherence generating power (CGP) with respect to a basis induced by the family. Specifically, the families of Hamiltonians were previously classified based on the scaling of the number $K$ of independent dynamically disconnected subspaces (called Krylov subspaces) with the system size\cite{moudgalya_hilbert_2022}. Each family of Hamiltonians is defined with respect to an algebra $\mathcal{A}$ generated by local terms that are used to compose the Hamiltonians. The generalized symmetries of the system are captured by the commutant algebra $\mathcal{A}^\prime$ and $K$ coincides with the dimension of the maximally Abelian subalgebra of $\mathcal{A}^\prime$. The Krylov subspaces are described by an algebra-induced Hilbert space decomposition, which also specifies a relevant basis. Our main result was then the demonstration that the maximum CGP (with respect to this relevant basis) of such a family of Hamiltonians is exactly related to $K$, hence its scaling with the system dimension is precisely dependent on the classification of the system. This gives an exact quantitative implementation of the intuitive connection between (generalized) symmetries and coherence generation. A principal example is the situation of Hilbert space fragmentation, in which case $K$ scales exponentially with the system size, leading to a substantially lower upper bound  for the CGP in finite dimensions. 
\par In order to further investigate the above observations, we numerically simulated different families of 1D spin-chain models and computed long time averages of their CGP with respect to the basis induced by the algebra of each family. We observed that for the fragmented systems, fermionic $t-J_z$ and spin-1 Temperley-Lieb (TL) models, the CGP follows closely the exponential behavior of the bound induced by the exponential number of Krylov subspaces. The particularly significant agreement in the TL case was connected with the previously observed Krylov restricted thermalization in some of its large Krylov subspaces in the thermodynamic limit. In contrast, in the case of the integrable spin-1/2 XXZ model the CGP greatly deviates from the bound set by the linear number of Krylov subspaces. Naturally, the above picture ties into the fact that the maximum (or close to the maximum) CGP is expected from systems that exhibit generic features inside sufficiently large Krylov subspaces.
\par The above observation is made precise by allowing the unitary evolution to be generated by any unitary in $\mathcal{A}$ and performing the Haar average of the CGP over all unitaries in $\mathcal{A}$. We show that the scaling of this average value with the system size depends exactly on the classification of the model in terms of the system size scaling of $K$. In addition, Levy's lemma ensures that the Haar average will be typical for sufficiently large Krylov subspaces.
\par A natural question one may ask is if there are more information-theoretic quantities that can be linked (in an exact fashion) with the classification of models induced by their symmetry algebra. Moreover, it would be interesting to further investigate the conditions under which families of models exhibit CGP close to the bound, as well as derive explicit connections with their ergodic and integrability properties inside the various Krylov subspaces.
\section{Acknowledgments}
 The Authors acknowledge discussions with E. Dallas. This research was (partially) sponsored by the Army Research Office and was accomplished under Grant Number W911NF-20-1-0075. The views and conclusions contained in this document are those of the authors and should not be interpreted as representing the official policies, either expressed or implied, of the Army Research Office or the U.S. Government. The U.S. Government is authorized to reproduce and distribute reprints for Government purposes notwithstanding any copyright notation herein.

\bibliographystyle{apsrev4-1}
\bibliography{main_CGP__fragmentation}

\onecolumngrid
\newpage
\appendix 
\section{Supplemental Material} \label{appendix}
\subsection{Proof of \autoref{prop_cgp}}
(i) Using \cref{eq_unitary} in \cref{eq_cgp_def} for the basis $B_\mathcal{A} = \{ \ket{\phi_J} \otimes \ket{\psi_J} \; \vert \; J=1,\dots,d_\mathcal{Z}, \; \phi_J = 1,\dots ,n_J, \; \psi_J = 1, \dots , d_J \}$ we get
\begin{equation} \label{eq_cgp_deriv}
\begin{split}
C_{B_\mathcal{A}}(U) &= 1-\frac{1}{d} \sum_{\phi_J , \phi^\prime_{J^\prime}} \sum_{J,J^\prime}\sum_{\psi_J , \psi^\prime_{J^\prime}} \lvert \bra{\phi_J}\otimes\bra{\psi_J} \oplus_{\tilde{J}} \mathds{1}_{n_{\tilde{J}}} \otimes U_{\tilde{J}} \ket{\phi^\prime_{J^\prime}} \otimes \ket{\psi^\prime_{J^\prime}}\rvert^4 \\
&= 1- \frac{1}{d} \sum_{\phi_J , \phi^\prime_{J^\prime}} \sum_{J,J'}\sum_{\psi_J , \psi^\prime_{J^\prime}} \delta_{J\tilde{J}}\delta_{J^\prime \tilde{J}} \delta_{\phi_J \phi^\prime_{J^\prime}} \lvert \mel{\psi_J}{U_{\tilde{J}}}{\psi^\prime_{J^\prime}}\rvert^4 = 1-\frac{1}{d} \sum_{J,\phi_J,\psi_J,\psi^\prime_J} \lvert \mel{\psi_J}{U_{J}}{\psi^\prime_J}\rvert^4 \\
&= 1-\frac{1}{d} \sum_{J,\psi_J,\psi^\prime_J} n_J \; \lvert \mel{\psi_J}{U_{J}}{\psi^\prime_J}\rvert^4
\end{split}
\end{equation}
(ii) Let $X\in \mathcal{L}(\mathbb{C}^{d_J})$, then since $\left\lVert X - \frac{\Tr X}{d_J} \mathds{1}_{d_J} \right\rVert_2^2 \geq 0$ we have
\begin{equation} \label{eq_norm_tr}
\lVert X \rVert_2^2 \geq \frac{\left\lvert \Tr X \right\rvert^2}{d_J} \quad  \forall \; X\in\mathcal{L}(\mathbb{C}^{d_J})
\end{equation}
Using the above identity for $X= \sum_{\psi_J} \lvert \mel{\psi_J}{U_J}{\psi_J^\prime} \rvert^2 \ket{\psi_J}\bra{\psi_J}$ we get
\begin{equation} \label{eq_ineq}
\sum_{\psi_J} \lvert \mel{\psi_J}{U_J}{\psi_J^\prime} \rvert^4 \geq \frac{\sum_{\psi_J} \lvert \mel{\psi_J}{U_J}{\psi_J^\prime} \rvert^2}{d_J} = \frac{1}{d_J}
\end{equation}
Using \cref{eq_ineq} in \cref{eq_cgp_deriv} we get
\begin{equation} \label{eq_max_deriv}
C_{B_\mathcal{A}}(U) \leq 1- \frac{1}{d} \sum_{J,\psi_J^\prime} n_J \frac{1}{d_J} = 1-\frac{1}{d} \sum_J d_J \; \frac{n_J}{d_J} = 1- \frac{1}{d} \sum_J n_J = 1-\frac{1}{d} K
\end{equation}
and clearly the maximum is achieved when $\lvert \mel{\psi_J}{U_J}{\psi_J^\prime} \rvert = d_J^{-1/2} \; \forall J$, i.e. when $U_J$ is mutually unbiased with respect to the basis $\{ \ket{\psi_J} \}$ of $\mathbb{C}^{d_J}$.
\subsection{Proof of \autoref{prop_haar}}
Due to \cref{eq_unitary}, taking the Haar average over all $U\in \mathcal{A}$ corresponds to taking the Haar average over the unitaries in the subsystems $\mathbb{C}^{d_J}$. Defining $\Pi_{\psi_J} = \ket{\psi_J}\bra{\psi_J}$ we can rewrite \cref{eq_cgp} as
\begin{equation} \label{eq_cgp_rewrite}
\begin{split}
C_{B_\mathcal{A}} (U) &= 1- \frac{1}{d} \sum_{J,\psi_J,\psi^\prime_J} n_J \; \left(\Tr [\Pi_{\psi_J} U_J \Pi_{\psi^\prime_J} U_J^\dagger ] \right)^2 = 1-\frac{1}{d} \sum_{J,\psi_J,\psi^\prime_J} n_J \; \Tr [(\Pi_{\psi_J} U_J \Pi_{\psi^\prime_J} U_J^\dagger \otimes \Pi_{\psi_J} U_J \Pi_{\psi^\prime_J} U_J^\dagger)] \\
&= 1-\frac{1}{d} \sum_{J,\psi_J, \psi^\prime_J} n_J \; \Tr [\Pi_{\psi_J}^{\otimes 2} U_J^{\otimes 2} \Pi_{\psi^\prime_J}^{\otimes 2} {U_J^\dagger}^{\otimes 2}]
\end{split}
\end{equation}
By Schur-Weyl duality the commutant of the algebra $\mathcal{M}_J$ generated by $\{ M_J^{\otimes 2} \, | \, M_J \in \mathcal{L}(\mathbb{C}^{d_J}) \}$ is $\mathcal{M}_J^\prime=\mathbf{C}S_2$, where $S_2=\{\mathds{1}, \, S\}$ is the symmetric group over the copies in $\mathbb{C}^{d_J} \otimes \mathbb{C}^{d_J}$ \cite{goodman_symmetry_2009}. Since we can always find a unitary basis of $\mathcal{L}(\mathbb{C}^{d_J})$, it follows that $\mathcal{M}_J$ is equivalently generated by $\{ U_J^{\otimes 2} \, | \, U_J \in \mathcal{L} (\mathbb{C}^{d_J} ), \, U_J\,U_J^\dagger = \mathds{1}_{d_J} \}$. Also, note that $\mathbb{P}_{\mathcal{M}_J^\prime} [\bullet ] := \overline{U_J^{\otimes 2} [\bullet ] {U_J^\dagger}^{\otimes 2}}^{\, {U_J}}$ is an orthogonal projector on $\mathcal{M}_J^\prime$ \footnote{It is not hard to check that $\mathbb{P}_{\mathcal{M}_J^\prime}^\dagger = \mathbb{P}_{\mathcal{M}_J^\prime}$, $\mathbb{P}_{\mathcal{M}_J^\prime}^2=\mathbb{P}_{\mathcal{M}_J^\prime}$, $\{\mathbb{P}_{\mathcal{M}_J^\prime}(M_J)\, | \, M_J \in \mathcal{L}({\mathbb{C}^{d_J}}^{\otimes 2})\} \subseteq \mathcal{M}_J^\prime$, $\mathbb{P}_{\mathcal{M}_J^\prime}(\mathds{1})=\mathds{1}$, $\mathbb{P}_{\mathcal{M}_J^\prime}(S)=S$.}. So, we can express $\mathbb{P}_{\mathcal{M}_J^\prime}$ in terms of the orthonormal basis $\left\{ \frac{\mathds{1} +S}{\sqrt{2d_J(d_J+1)}},\frac{\mathds{1} -S}{\sqrt{2d_J(d_J-1)}} \right\}$ of $\mathbf{C}S_2$:
\begin{align}
\mathbb{P}_{\mathcal{M}_J^\prime} [\bullet]= \overline{U_J^{\otimes 2} [\bullet ] {U_J^\dagger}^{\otimes 2}}^{\, {U_J}} = \sum_{\eta = \pm 1} \frac{\mathds{1} + \eta \, S }{2d_J (d_J+ \eta )} \langle \mathds{1} + \eta \, S , \bullet \rangle \label{eq_schur}
\end{align}
Taking the Haar average over the unitaries in $\mathbb{C}^{d_J}$ in \cref{eq_cgp_rewrite} and using \cref{eq_schur} we get
\begin{equation} \label{cgp_haar_deriv}
\begin{split}
\overline{C_{B_\mathcal{A}}(U)}^{U \in \mathcal{A}} &= 1- \frac{1}{d} \sum_{J,\psi_J, \psi^\prime_J} n_J \; \Tr [\Pi_{\psi_J}^{\otimes 2} \sum_{\eta = \pm 1} \frac{\mathds{1} + \eta \, S }{2d_J (d_J+ \eta )} \langle \mathds{1} + \eta \, S , \Pi_{\psi^\prime_J} \rangle ] = 1 -\frac{1}{d} \sum_{J,\psi_J,\psi^\prime_J} n_J \; \Tr [\Pi_{\psi_J}^{\otimes 2} \frac{\mathds{1}+S}{d_J(d_J+1)}] \\
&=1-\frac{1}{d} \sum_{J,\psi_J,\psi^\prime_J} n_J \frac{2}{d_J (d_J+1)} = 1-\frac{1}{d} \sum_J \frac{2 d_J}{d_J+1} n_J
\end{split}
\end{equation}
\subsection{Typicality inside sufficiently large Krylov subspaces} \label{append_typicality}
This follows from Proposition 4 of \cite{zanardi_coherence-generating_2017}. The key ingredient is Levy's lemma for Haar distributed unitaries in $\mathbb{C}^{d_J}$. The operator norms used are the Schatten $k$-norms\cite{bhatia_symmetric_1997} defined as $\lVert X \rVert_k := \left(\sum_i s_i^k \right)^{1/k}$, where $\{ s_i \}_i$ are the singular values of $X$. For $k\rightarrow \infty$, $\lVert X \rVert_\infty = \max_i \{s_i\}$ is the usual operator norm. 
\begin{lemma}[Levy's lemma] \label{lemma_levy}
If $X: \; U(d^J) \mapsto \mathbb{R}$ is a Lipschitz continuous function of constant $\lambda$, i.e. $\lvert X(U_J) - X(V_J) \rvert \leq \lambda \lVert X(U_J) - X(V_J) \rVert_2$, then 
\begin{equation} \label{eq_levy}
\Pr  (\left\lvert X(U_J) - \overline{X(U_J)}^{U_J \in \mathbb{C}^{d_J}} \right\rvert\geq \epsilon ) \leq e^{-\frac{d_J \; \epsilon^2}{4 \lambda^2}}
\end{equation}
\end{lemma}
We shall also need known norm inequalities such as
\begin{align}
&\lvert \Tr [AB] \rvert \leq \lVert A \rVert_1 \lVert B \rVert_\infty \label{eq_ineq1}\\
&\text{If }\mathcal{T}\text{ is a positive, trace-preserving map, }\lVert \mathcal{T}(X) \rVert_1 \leq \lVert X \rVert_1 \label{eq_ineq2}\\
&\frac{1}{2}\lVert \ket{\psi}\bra{\psi} - \ket{\phi}\bra{\phi} \rVert_1 \leq \lVert \ket{\psi} - \ket{\phi} \rVert \leq 2 \label{eq_ineq3}
\end{align}
Choose $U=\oplus_J \mathds{1}_{n_J} \otimes U_J, \; V=\oplus_J \mathds{1}_{n_J} \otimes V_J \in \mathcal{A}$. Note that using the swap trick $\Tr[AB] = \Tr[ S A\otimes B]$, we can rewrite
\begin{equation}
C_{B_\mathcal{A}}(U) = 1- \frac{1}{d} \sum_J n_J \sum_{\psi_J,\psi^\prime_J}\lvert \mel{\psi_J}{U_J}{\psi^\prime_J} \rvert^4 = 1- \frac{1}{d} \sum_J n_J \sum_{\psi^\prime_J} \Tr[ S_J \left(\mathcal{D}_J\mathcal{U}_J (\Pi_{\psi^\prime_J})\right)^{\otimes 2}]
\end{equation}
where $D_J (\bullet ) = \sum_{\psi_J} \Pi_{\psi_J} \bullet \Pi_{\psi_J}$, $\mathcal{U}_J (\bullet )= U_J \bullet U_J^\dagger$ and $S_J$ is the swap in $\mathbb{C}^{d_J}\otimes\mathbb{C}^{d_J}$.  Now, let $X(U_J)=\frac{1}{d_J}\sum_{\psi^\prime_J} \Tr[ S_J \mathcal{D}_J^{\otimes 2}\mathcal{U}_J^{\otimes 2}(\Pi_{\psi^\prime_J}^{\otimes 2})]$. We have:
\begin{equation}
\begin{split}
\lvert X(U_J) - X(V_J) \rvert &= \frac{1}{d_J}\left\lvert \sum_{\psi^\prime_J} \Tr[ S_J \mathcal{D}_J^{\otimes 2} \left(\mathcal{U}_J^{\otimes 2} - \mathcal{V}_J^{\otimes 2}\right)\left(\Pi_{\psi^\prime_J}^{\otimes 2}\right)] \right\rvert \leq \\
&\leq \frac{1}{d_J} \sum_{\psi^\prime_J} \left\lvert  \Tr[ S_J \mathcal{D}_J^{\otimes 2} \left(\mathcal{U}_J^{\otimes 2} - \mathcal{V}_J^{\otimes 2}\right)\left(\Pi_{\psi^\prime_J}^{\otimes 2}\right)] \right\rvert\leq\\
&\leq \frac{1}{d_J} \sum_{\psi^\prime_J} \left\lVert S\right\rVert_\infty \left\lVert \mathcal{D}_J^{\otimes 2} \left(\mathcal{U}_J^{\otimes 2} - \mathcal{V}_J^{\otimes 2}\right)\left(\Pi_{\psi^\prime_J}^{\otimes 2}\right) \right\rVert_1 \leq  \\
&\leq \frac{1}{d_J} \sum_{\psi^\prime_J} \left\lVert \mathcal{U}_J^{\otimes 2}(\Pi_{\psi^\prime_J}^{\otimes 2}) - \mathcal{V}_J^{\otimes 2}(\Pi_{\psi^\prime_J}^{\otimes 2}) \right\rVert \leq \\
&\leq \frac{1}{d_J} \sum_{\psi^\prime_J} 2 \left\lVert U_J^{\otimes 2} \ket{\psi^\prime_J}^{\otimes 2} - V_J^{\otimes 2} \ket{\psi^\prime_J}^{\otimes 2} \right\rVert \leq \frac{1}{d_J} \sum_{\psi^\prime_J} 2 \left\lVert U_J^{\otimes 2} - V_J^{\otimes 2} \right\rVert_\infty = 2 \left\lVert U_J^{\otimes 2} - V_J^{\otimes 2} \right\rVert_\infty =\\
&= 2 \lVert \mathds{1} - {U_J^\dagger}^{\otimes 2} V_J^{\otimes 2} \rVert_\infty
\end{split}
\end{equation}
where to go from line 1 to line 2 we used triangle inequality, from line 2 to line 3 \cref{eq_ineq1} with $A=\mathcal{D}_J^{\otimes 2} \left(\mathcal{U}_J^{\otimes 2} - \mathcal{V}_J^{\otimes 2}\right)\left(\Pi_{\psi^\prime_J}^{\otimes 2}\right)$, $B=S$, from line 3 to line 4 the fact that $\lVert S \rVert_\infty = 1$ and \cref{eq_ineq2} with $\mathcal{T}=\mathcal{D}_J$, from line 4 to line 5 \cref{eq_ineq3} with $\ket{\psi} = U_J^{\otimes 2} \ket{\psi_J^\prime }^{\otimes 2}$, $\ket{\phi} = V_J^{\otimes 2} \ket{\psi_J^\prime}^{\otimes 2}$ and in line 5 the definition of the operator norm.
Let us define $M:= \mathds{1}_{d_J} - U_J^\dagger V_J$. Then:
\begin{equation}
\begin{split}
\lvert X(U_J) - X(V_J) \rvert &\leq 2 \left\lVert \mathds{1} - (\mathds{1}+M\otimes M - \mathds{1}_{d_J} \otimes M - M \otimes \mathds{1}_{d_J} )\right\rVert_\infty \leq \\
&\leq 2 \left( 2\lVert M \rVert_\infty + \lVert M \rVert_\infty^2 \right) \leq 8 \lVert M \rVert_\infty = 8 \lVert U_J - V_J \rVert_\infty \leq 8 \lVert U_J -V_J \rVert_2
\end{split}
\end{equation}
where we used that $\lVert M \rVert_\infty \leq 2$. So, $X:U(\mathbb{C}^{d_J}) \mapsto \mathbb{R}$ is Lipschitz continuous with constant $\lambda =8$ and the result follows from \cref{lemma_levy}.

\subsection{System dimension scalings}\label{dimensionality_details}
The scaling of $\overline{f_{B_{t-J_z}}(U_t)}^t$ and $\overline{f_{B_{TL}}(U_t)}^t$ in \cref{spin_chain_sect} is found by finding the best exponential-law ($f(x)=A\; x^B$) fit of the numerical data. Specifically, we find $\overline{f_{B_{t-J_z}}(U_t)}^t \sim d^{\lambda}$ with $\lambda=0.8340\pm 0.0023$ and $RMSE=2.989$ and $\overline{f_{B_{TL}}(U_t)}^t\sim d^\lambda$ with $\lambda = 0.9468 \pm 0.0016$ and $RMSE=2.192$.
\par The scalings of $K_{t-J_z}$ and $K_{TL}$ follow directly from the expressions $K_{t-J_z}=2^{L+1}-1$ and $K_{TL}=\frac{q^{-L} \left(q^{L+2}-1\right)^2}{\left( q^2 -1 \right)^2}$, where in both cases $L=\log_3 d$. Explicitly:
\begin{equation*}
\begin{split}
    &K_{t-J_z}=2^{L+1}-1=2^{\log_3 d +1}-1 \sim 2^{\log_3 d} = d^{\log_3 2}\\
    &K_{TL}=\frac{q^{-L} \left(q^{L+2}-1\right)^2}{\left( q^2 -1 \right)^2}\sim q^L = q^{\log_3 d} = d^{\log_3 q}
\end{split}
\end{equation*}

\end{document}